# On constrained second derivatives

Tamás Gál

Quantum Theory Project, Department of Physics, University of Florida,
Gainesville, Florida 32611, USA

**Abstract:** The question of defining unique, generally applicable constrained second, and higher-order, derivatives is investigated. It is shown that second-order constrained derivatives obtained via two successive constrained differentiations provide a proper tool for the determination of the characters of stationary points under constraints. As an illustration, a finite-dimensional example is considered that has previously been used as an illustration for other stationary-point analysis methods.

## I. Introduction

Functional derivatives [1] play an essential role in the description of physical phenomena. However, the general proper account for constraints when functional derivatives of any order are used is yet an unsolved question; this had been the case until very recently even in the analysis of stationary points (see e.g. [2]). Recently, it has been shown [3] that second-order constrained functional derivatives defined as

$$\frac{\delta^2 A[\rho]}{\delta_C \rho(x) \delta_C \rho(x')} := \frac{\delta^2 A[\rho_C[\rho]]}{\delta \rho(x) \delta \rho(x')}\bigg|_{\rho=\rho_C} \tag{1}$$

provide a proper tool to carry out analyses of the characters of stationary points in the presence of constraints on the functional variable $\rho(x)$. In Eq.(1), $\rho_C[\rho]$ is a functional that projects the set of $\rho(x)$'s onto the set of $\rho_C(x)$'s (functions that satisfy the given constraint $C[\rho] = C$); that is, $\rho_C[\rho]$ gives an identity for $\rho_C(x)$'s, otherwise it delivers, for any $\rho(x)$, some corresponding $\rho_C(x)$. In fact, with the use of $\rho_C[\rho]$, constrained derivatives of any order $m$ can be defined:

$$\frac{\delta^m A[\rho]}{\delta_C \rho^m} := \frac{\delta^m A[\rho_C[\rho]]}{\delta \rho^m}\bigg|_{\rho=\rho_C}. \tag{2}$$

With these derivatives, then, the Taylor series expansion of a functional over a domain restricted by some constraint can be written as

$$A[\rho + \Delta_C \rho] = A[\rho] + \int \frac{\delta A[\rho]}{\delta_C \rho(x)} \Delta \rho(x) dx + \frac{1}{2!} \iint \frac{\delta^2 A[\rho]}{\delta_C \rho(x) \delta_C \rho(x')} \Delta \rho(x) \Delta \rho(x') dx dx'$$

$$+ \frac{1}{3!} \iiint \frac{\delta^3 A[\rho]}{\delta_C \rho(x) \delta_C \rho(x') \delta_C \rho(x'')} \Delta \rho(x) \Delta \rho(x') \Delta \rho(x'') dx dx' dx'' + ..., \tag{3}$$

where $\Delta_C \rho(x)$ is any increment of the functional variable $\rho(x)$ that satisfies the given constraint, i.e. $C[\rho + \Delta_C \rho] = C[\rho]$. As can be seen, the constrained derivatives free the variations of $\rho(x)$ from the constraint in the Taylor expansion, incorporating all effects of the constraint, which is their essential feature that makes it possible to substitute normal functional derivatives by constrained ones in applications of derivatives in the presence of constraints. Constrained second derivatives have been successfully applied in a stability analysis of droplet growth in supercooled vapors [4] and in the determination of the character of quantum mechanical eigenstates as stationary points of the energy expectation value [5].

The basis of these applications is the fact [3] that the eigenvalues of Eq.(1), i.e. the solutions $\lambda$ of the eigenvalue equation

$$\int \frac{\delta^2 A[\rho]}{\delta_C\rho(x)\delta_C\rho(x')} \Delta\rho(x')dx' = \lambda\Delta\rho(x) , \qquad (4)$$

determine whether a given stationary point is a local minimum/maximum or a saddle point, according to that the $\lambda$'s are all positive/negative or both (provided that zero is not an accumulation point of the $\lambda$'s).

In [3], no concrete form for $\rho_C[\rho]$ was proposed; in fact, as has been shown, one may choose any $\rho_C[\rho]$ with the above properties to obtain proper constrained derivatives for a stationary point analysis. In general, first-order constrained derivatives must have the form [6]

$$\frac{\delta A[\rho]}{\delta_C\rho(x)} = \frac{\delta A[\rho]}{\delta\rho(x)} - \frac{\delta C[\rho]}{\delta\rho(x)} \int \left( u(x') \bigg/ \frac{\delta C[\rho]}{\delta\rho(x')} \right) \frac{\delta A[\rho]}{\delta\rho(x')} dx' , \qquad (5)$$

with $u(x')$ a function that integrates to one; more specifically [3], if Eq.(5) is originated from some $\rho_C[\rho]$,

$$u(x) = \frac{\delta C[\rho]}{\delta\rho(x)} \frac{\partial \rho_C(x)[\rho]}{\partial C} . \qquad (6)$$

More recently, it has been shown [5] that with the choice

$$u(x) = \left( \frac{\delta C[\rho]}{\delta\rho(x)} \right)^2 \bigg/ \int \left( \frac{\delta C[\rho]}{\delta\rho(x')} \right)^2 dx' \qquad (7)$$

in Eq.(5), the eigenvalues of Eq.(4), with

$$\frac{\delta^2 A[\rho]}{\delta_C\rho(x)\delta_C\rho(x')} = \iint \frac{\delta\rho(x'')}{\delta_C\rho(x)} \left( \frac{\delta^2 A[\rho]}{\delta\rho(x'')\delta\rho(x''')} - \mu \frac{\delta^2 C[\rho]}{\delta\rho(x'')\delta\rho(x''')} \right) \frac{\delta\rho(x''')}{\delta_C\rho(x')} dx''dx''' , \qquad (8)$$

determine not only the character of the given constrained stationary point, but even the Morse index of that point; namely, the maximum number of orthogonal eigenfunctions with negative eigenvalues will deliver that index. $\frac{\delta\rho(x')}{\delta_C\rho(x)}$ in Eq.(8) is the constrained first derivative of $\rho(x)$ calculated according to Eq.(5) with Eq.(7) inserted, while $\mu$ is the Lagrange multiplier corresponding to the considered constrained stationary point, i.e. $\mu = \frac{\delta A[\rho]}{\delta\rho(x)} \bigg/ \frac{\delta C[\rho]}{\delta\rho(x)}$.

The constrained first derivative $\frac{\delta A[\rho]}{\delta_C\rho(x)}$ obtained with the use of Eq.(7),

$$\frac{\delta A[\rho]}{\delta_C\rho(x)} = \frac{\delta A[\rho]}{\delta\rho(x)} - \frac{\delta C[\rho]}{\delta\rho(x)} \frac{1}{Q[\rho]} \int \frac{\delta C[\rho]}{\delta\rho(x')} \frac{\delta A[\rho]}{\delta\rho(x')} dx' , \qquad (9)$$

with $Q[\rho] \doteq \int \left( \frac{\delta C[\rho]}{\delta \rho(x)} \right)^2 dx$, is nothing else than the orthogonal projection of the full derivative $\frac{\delta A[\rho]}{\delta \rho(x)}$ onto the constraint domain of $\rho(x)$'s. This is because $\frac{\delta C[\rho]}{\delta \rho(x)}$ is a vector that is orthogonal to the constraint domain; consequently, the second term of $\frac{\delta A[\rho]}{\delta_C \rho(x)}$ of Eq.(9) is a component of $\frac{\delta A[\rho]}{\delta \rho(x)}$ that is orthogonal to the constraint domain (the $Q^{-1/2}$ factor ensures the normalization of the vector $\frac{\delta C[\rho]}{\delta \rho(x)}$). We note that in a general Banach space of real functions, orthogonality of functions as defined by $\int f(x)g(x)dx = 0$ is not unique, in the sense that there are several functions (differing by more than a constant factor) that are orthogonal to a given domain $D[\rho] = D$ – consequently, $\frac{\delta C[\rho]}{\delta \rho(x)}$ is only one choice of functions orthogonal to the constraint domain, implying that the orthogonal projection of $\frac{\delta A[\rho]}{\delta \rho(x)}$ onto the constraint domain is not unique either. However, the choice Eq.(7), i.e. the projection Eq.(9) of $\frac{\delta A[\rho]}{\delta \rho(x)}$, is still special, considering its advantage with respect to the determination of the Morse index of stationary points [5]. (As seen in [5], in the Hilbert space of complex-valued functions with scalar product defined as $\int f^*(x)g(x)dx$, the problem investigated here did not arise: In the case considered there, it was possible to find a $\rho_C[\rho]$ that gives the orthogonal projection of a full derivative.) In the finite-dimensional case, i.e. in the case of multi-variable functions $a(\rho_1,...,\rho_n)$, instead of functionals $A[\rho]$, orthogonality in the variable domain becomes unique. The finite-dimensional version of Eq.(9) can be given as

$$\frac{\partial a(\bar{\rho})}{\partial_C \rho_i} = \frac{\partial a(\bar{\rho})}{\partial \rho_i} - \frac{\partial c(\bar{\rho})}{\partial \rho_i} \frac{\nabla c(\bar{\rho}) \cdot \nabla a(\bar{\rho})}{|\nabla c(\bar{\rho})|^2} \quad , \tag{10a}$$

or compactly,

$$\nabla_C a(\bar{\rho}) = \nabla a(\bar{\rho}) - \nabla c(\bar{\rho}) \frac{\nabla c(\bar{\rho}) \cdot \nabla a(\bar{\rho})}{|\nabla c(\bar{\rho})|^2} \quad , \tag{10b}$$

where $\nabla$ denotes the *n*-dimensional gradient $\left( \frac{\partial}{\partial \rho_1},..., \frac{\partial}{\partial \rho_n} \right)$, acting on scalar functions of vector variables $\bar{\rho}$ of *n* dimensions. The choice Eq.(7) is then really appealing since when e.g. a surface $a(\rho_1,\rho_2)$ needs to be differentiated over a curve in the domain of $(\rho_1,\rho_2)$'s, determined by a constraint $c(\rho_1,\rho_2)=c$, Eq.(10) gives back just the directional derivative times the unit vector pointing in the direction of the tangential of $c(\rho_1,\rho_2)$ at the considered point.

The above choice for constrained derivatives, however, raises some problems in the way of their general use. For instance, in the case of perhaps the simplest constraint, $\int \rho(x)\,dx = N$, Eq.(9) gives

$$\frac{\delta A[\rho]}{\delta_N \rho(x)} = \frac{\delta A[\rho]}{\delta \rho(x)} - \frac{1}{\int dx''} \int \frac{\delta A[\rho]}{\delta \rho(x')}\,dx' , \qquad (11)$$

which is problematic if the integration goes over an infinite-size domain. Hence, in the density functional theory of electronic systems [7], e.g., Eq.(11) is not useful. Originally, another choice of $u(x)$ was proposed in [8]: $u(x) = \rho(x)\big/\int \rho(x')dx'$, which emerges from $\rho_N[\rho] = \frac{N}{\int \rho(x')dx'}\rho(x)$, and yields a constrained derivative that gives back the full derivative itself for degree-zero homogeneous functionals [6]. Since density functional theory is based on variational principles, i.e. is centered on stationary points, the concrete choice of $\rho_N[\rho]$ is irrelevant there. (Further, in the case of normalization conservation of wave-functions [5], the degree-zero homogeneous choice of $\rho_C[\rho]$ is the one that yields the orthogonal projection of the full derivative as the constrained derivative.)

There is a more general, mathematical, problem with the choice Eq.(7). In general, Eq.(7) cannot be derived from a $\rho_C[\rho]$ through Eq.(6) – except for linear constraints, i.e. constraints of the form

$$\int g(x)\rho(x)dx = L . \qquad (12)$$

For linear constraints, with $\frac{\delta L[\rho]}{\delta \rho(x)} = g(x)$,

$$\rho_L[\rho] = \rho(x) - \frac{g(x)}{\int (g(x''))^2 dx''}\left(\int g(x')\rho(x')dx' - L\right) \qquad (13)$$

will give Eq.(9), as can be checked simply by differentiating Eq.(13) with respect to $L$ (while keeping $\rho(x)$ fixed) and then multiplying the result by $g(x)$, according to Eq.(6), or directly by using Eq.(2) ($m$=1). For non-linear constraints,

$$\rho_C[\rho] = \rho(x) - \frac{\frac{\delta C[\rho]}{\delta \rho(x)}}{\int \left(\frac{\delta C[\rho]}{\delta \rho(x')}\right)^2 dx'} (C[\rho] - C) \qquad (14)$$

would similarly give Eq.(9), but the problem is that Eq.(14) does not satisfy the defining criteria of a $\rho_C[\rho]$ – concretely, it does not satisfy $C[\rho] = C$. This is indeed a problem since due to this, Eq.(14) will not yield proper higher-order constrained derivatives that can be used e.g. in stationary point analyses. This was the reason to write Eq.(8) instead of referring to Eq.(1). However, although constrained second derivatives defined by Eq.(1) give Eq.(8) at stationary points [3] (which is the cornerstone of their applicability in stationary point analysis), Eq.(8) cannot be considered as a general definition of constrained second derivatives, applicable for non-stationary situations. Giving a definition of higher-order constrained derivatives that is consistent with Eq.(9) would be highly desirable for general physical applications; therefore, we will examine this question in this study.

## II. On a general definition of constrained second derivatives

The reason for that Eq.(14) yields proper first-order constrained derivatives is that it satisfies the constraint up to first order. By this, we mean that an infinitesimal increment of a $\rho_C(x)$ to some $\rho_C[\rho](x)$ given by Eq.(14) satisfies

$$\int \frac{\delta C[\rho]}{\delta \rho(x)} \Delta_{C'} \rho(x) = 0 \ . \qquad (15)$$

(Earlier, we denoted an increment of a $\rho_C(x)$ to some $\tilde{\rho}_C(x)$ that also satisfies the constraint by $\Delta_C \rho(x)$, and now $\Delta_{C'} \rho(x)$, $\Delta_{C''} \rho(x)$, etc. denote increments satisfying the given constraint up to first order, second order, etc., respectively.) More precisely speaking, if we insert the Taylor expansion

$$\rho_C[\rho](x) - \rho_C(x) = \int \frac{\delta \rho_C[\rho_C](x)}{\delta \rho(x')} \Delta \rho(x') \, dx' + \frac{1}{2} \int \frac{\delta^2 \rho_C[\rho_C](x)}{\delta \rho(x') \delta \rho(x'')} \Delta \rho(x') \Delta \rho(x'') \, dx' dx'' + .... \qquad (16)$$

of $\rho_C[\rho](x)$ around some $\rho_C(x)$, with $\Delta \rho(x)$ denoting $\rho(x) - \rho_C(x)$, in the Taylor expansion

$$C[\rho] - C = \int \frac{\delta C[\rho_C]}{\delta \rho(x)} \Delta \rho(x) \, dx + \frac{1}{2!} \iint \frac{\delta^2 C[\rho_C]}{\delta \rho(x) \delta \rho(x')} \Delta \rho(x) \Delta \rho(x') \, dx dx' + ... \qquad (17)$$

of $C[\rho]$ around $\rho_C(x)$ (i.e., substitute $\rho_C[\rho](x) - \rho_C(x)$ of Eq.(16) for $\Delta \rho(x)$ in Eq.(17)), we will get Eq.(15) after neglecting terms of order higher than one in $\Delta \rho(x)$. This can be checked

readily – for this, note that we need to check the fulfillment of Eq.(15) only at $\rho_C(x)$; thus, the derivative of the factor of $(C[\rho]-C)$ in Eq.(14) will cancel in the process. This results in

$$\iint \frac{\delta C[\rho]}{\delta \rho(x)} \frac{\delta \rho_C[\rho](x)}{\delta \rho(x')}\bigg|_{\rho=\rho_C} dx \Delta\rho(x')\, dx' = 0 \tag{18}$$

for any $\Delta\rho(x)$. Notice that Eq.(18) implies

$$\int \frac{\delta C[\rho]}{\delta \rho(x)} \frac{\delta \rho(x)}{\delta_C \rho(x')}\, dx = 0 \tag{19}$$

(with the use of Eq.(2)) or, in other words,

$$\frac{\delta C[\rho]}{\delta_C \rho(x')} = 0 \;. \tag{20}$$

Since

$$\frac{\delta \rho(x)}{\delta_C \rho(x')} = \delta(x-x') - \frac{1}{Q} \frac{\delta C[\rho]}{\delta \rho(x')} \frac{\delta C[\rho]}{\delta \rho(x)} \tag{21}$$

is symmetric in its $x$ arguments, integration, in $x'$, of Eq.(19) multiplied by $\dfrac{\delta A[\rho]}{\delta \rho(x')}$ gives

$$\int \frac{\delta C[\rho]}{\delta \rho(x)} \frac{\delta A[\rho]}{\delta_C \rho(x)}\, dx = 0 \;. \tag{22}$$

Eq.(22) is just what we expect from a constrained first derivative: It has zero component that is orthogonal to the constraint domain.

Thus, to obtain constrained first derivatives, we do not need to require $\rho_C[\rho]$ to satisfy the given constraint, but it is enough if $\rho_C[\rho]$ satisfies the constraint up to first order. (Then, the proper choice for $\rho_C[\rho]$ is given by Eq.(14).) This observation naturally leads to the idea that to obtain constrained second derivatives, it is satisfactory to find a $\rho_C[\rho]$ that satisfies the given constraint up to second order. That is, $\rho_C[\rho](x) - \rho_C(x)$ should satisfy

$$\int \frac{\delta C[\rho]}{\delta \rho(x)} \Delta_{C''}\rho(x)\, dx + \frac{1}{2}\iint \frac{\delta^2 C[\rho]}{\delta \rho(x)\delta \rho(x')} \Delta_{C''}\rho(x) \Delta_{C''}\rho(x')\, dx dx' = 0 \tag{23}$$

for infinitesimal increments of $\rho_C(x)$ to a general $\rho(x)$ that are negligible in third or higher order. Inserting Eq.(16) in Eq.(17), we obtain up to second order

$$\iiint \frac{\delta C[\rho_C]}{\delta \rho(x'')} \frac{\delta^2 \rho_C[\rho_C](x)}{\delta \rho(x) \delta \rho(x')} \Delta\rho(x)\Delta\rho(x')\, dx dx' dx'' + \iiiint \frac{\delta^2 C[\rho_C]}{\delta \rho(x'')\delta \rho(x''')} \frac{\delta \rho_C[\rho_C](x'')}{\delta \rho(x)} \frac{\delta \rho_C[\rho_C](x''')}{\delta \rho(x')} \Delta\rho(x)\Delta\rho(x')\, dx dx' dx'' dx''' = 0, \tag{24}$$

the condition to be satisfied by the searched $\rho_C[\rho]$. Since Eq.(24) is to hold for any infinitesimal $\Delta\rho(x)$, it gives

$$\int \frac{\delta C[\rho_C]}{\delta\rho(x'')} \frac{\delta^2 \rho_C[\rho_C](x'')}{\delta\rho(x)\delta\rho(x')} dx'' + \iint \frac{\delta^2 C[\rho_C]}{\delta\rho(x'')\delta\rho(x''')} \frac{\delta\rho_C[\rho_C](x'')}{\delta\rho(x)} \frac{\delta\rho_C[\rho_C](x''')}{\delta\rho(x')} dx''dx''' = 0 \ . \qquad (25)$$

Note that in Eq.(24), the first-order term in $\Delta\rho(x)$ has been dropped, since it has to vanish separately, i.e. Eq.(18) will hold.

If we substitute Eq.(14) for $\rho_C[\rho]$ in Eq.(25), we find that a term $\frac{1}{Q^2} \frac{\delta C[\rho]}{\delta\rho(x)} \frac{\delta C[\rho]}{\delta\rho(x')} \iint \frac{\delta^2 C[\rho]}{\delta\rho(x'')\delta\rho(x''')} \frac{\delta C[\rho]}{\delta\rho(x'')} \frac{\delta C[\rho]}{\delta\rho(x''')} dx''dx'''$ remains to be cancelled. An opposite-sign term may come only from a second-order modification of Eq.(14), since we want to keep Eq.(14) as the first-order term in the searched $\rho_C[\rho]$, in order to not modify the constrained first derivative Eq.(9). An appropriate modification of Eq.(14) can then be e.g.

$$\rho_C[\rho] = \rho(x) - \frac{1}{Q[\rho]} \frac{\delta C[\rho]}{\delta\rho(x)} (C[\rho] - C) - \frac{1}{2(Q[\rho])^2} \int \frac{\delta^2 C[\rho]}{\delta\rho(x)\delta\rho(x')} \frac{\delta C[\rho]}{\delta\rho(x')} dx' (C[\rho] - C)^2 . \qquad (26)$$

Eq.(26) yields only a one-term contribution to the constrained second derivative obtained with Eq.(14), indeed, because of the cancellation of other terms in the first and second derivatives of Eq.(26), due to $C[\rho_C] = C$. For these calculations, note that

$$\frac{\delta Q[\rho]}{\delta\rho(x)} = 2\int \frac{\delta^2 C[\rho]}{\delta\rho(x)\delta\rho(x')} \frac{\delta C[\rho]}{\delta\rho(x')} dx' \ . \qquad (27)$$

Eq.(26) yields Eq.(9) as constrained first derivative, as it should. To obtain the constrained second derivative emerging from Eq.(26) via Eq.(1), it is worth applying the chain rule of differentiation in Eq.(1) as an intermediate step, which yields [3]

$$\frac{\delta^2 A[\rho]}{\delta_C\rho(x)\delta_C\rho(x')} = \iint \frac{\delta^2 A[\rho]}{\delta\rho(x'')\delta\rho(x''')} \frac{\delta\rho(x'')}{\delta_C\rho(x)} \frac{\delta\rho(x''')}{\delta_C\rho(x')} dx''dx''' + \int \frac{\delta A[\rho]}{\delta\rho(x'')} \frac{\delta^2\rho(x'')}{\delta_C\rho(x)\delta_C\rho(x')} dx'' \ . \qquad (28)$$

The constrained first derivative of $\rho(x)$ is given by Eq.(21), while the second derivative arises from Eq.(26) on the basis of Eq.(1) as

$$\frac{\delta^2\rho(x'')}{\delta_C\rho(x)\delta_C\rho(x')} = -\frac{1}{Q} \frac{\delta^2 C[\rho]}{\delta\rho(x'')\delta\rho(x')} \frac{\delta C[\rho]}{\delta\rho(x)} - \frac{1}{Q} \frac{\delta^2 C[\rho]}{\delta\rho(x'')\delta\rho(x)} \frac{\delta C[\rho]}{\delta\rho(x')} - \frac{1}{Q} \frac{\delta^2 C[\rho]}{\delta\rho(x)\delta\rho(x')} \frac{\delta C[\rho]}{\delta\rho(x'')}$$
$$+ \frac{2}{Q^2} \frac{\delta C[\rho]}{\delta\rho(x')} \frac{\delta C[\rho]}{\delta\rho(x'')} \int \frac{\delta^2 C[\rho]}{\delta\rho(x)\delta\rho(x''')} \frac{\delta C[\rho]}{\delta\rho(x''')} dx''' + \frac{2}{Q^2} \frac{\delta C[\rho]}{\delta\rho(x)} \frac{\delta C[\rho]}{\delta\rho(x'')} \int \frac{\delta^2 C[\rho]}{\delta\rho(x')\delta\rho(x''')} \frac{\delta C[\rho]}{\delta\rho(x''')} dx'''$$
$$- \frac{1}{Q^2} \frac{\delta C[\rho]}{\delta\rho(x)} \frac{\delta C[\rho]}{\delta\rho(x')} \int \frac{\delta^2 C[\rho]}{\delta\rho(x'')\delta\rho(x''')} \frac{\delta C[\rho]}{\delta\rho(x''')} dx''' \ . \qquad (29)$$

Inserting these expressions in Eq.(28) yields the constrained second derivative of a functional $A[\rho]$:

$$\frac{\delta^2 A[\rho]}{\delta_C\rho(x)\delta_C\rho(x')} = \frac{\delta^2 A[\rho]}{\delta\rho(x)\delta\rho(x')} - \frac{1}{Q}\frac{\delta C[\rho]}{\delta\rho(x)}\left(\int \frac{\delta C[\rho]}{\delta\rho(x'')}\frac{\delta^2 A[\rho]}{\delta\rho(x')\delta\rho(x'')}dx'' + \int \frac{\delta^2 C[\rho]}{\delta\rho(x')\delta\rho(x'')}\frac{\delta A[\rho]}{\delta\rho(x'')}dx''\right.$$

$$\left. - \frac{2}{Q}\int \frac{\delta C[\rho]}{\delta\rho(x'')}\frac{\delta^2 C[\rho]}{\delta\rho(x')\delta\rho(x'')}dx''\int \frac{\delta C[\rho]}{\delta\rho(x''')}\frac{\delta A[\rho]}{\delta\rho(x''')}dx'''\right)$$

$$-\frac{1}{Q}\frac{\delta C[\rho]}{\delta\rho(x')}\left(\int \frac{\delta C[\rho]}{\delta\rho(x'')}\frac{\delta^2 A[\rho]}{\delta\rho(x)\delta\rho(x'')}dx'' + \int \frac{\delta^2 C[\rho]}{\delta\rho(x)\delta\rho(x'')}\frac{\delta A[\rho]}{\delta\rho(x'')}dx''\right.$$

$$\left. - \frac{2}{Q}\int \frac{\delta C[\rho]}{\delta\rho(x'')}\frac{\delta^2 C[\rho]}{\delta\rho(x)\delta\rho(x'')}dx''\int \frac{\delta C[\rho]}{\delta\rho(x''')}\frac{\delta A[\rho]}{\delta\rho(x''')}dx'''\right)$$

$$+\frac{1}{Q^2}\frac{\delta C[\rho]}{\delta\rho(x)}\frac{\delta C[\rho]}{\delta\rho(x')}\left(\iint \frac{\delta C[\rho]}{\delta\rho(x'')}\frac{\delta C[\rho]}{\delta\rho(x''')}\frac{\delta^2 A[\rho]}{\delta\rho(x'')\delta\rho(x''')}dx''dx''' - \iint \frac{\delta^2 C[\rho]}{\delta\rho(x'')\delta\rho(x''')}\frac{\delta C[\rho]}{\delta\rho(x''')}\frac{\delta A[\rho]}{\delta\rho(x'')}dx''dx'''\right)$$

$$-\frac{1}{Q}\frac{\delta^2 C[\rho]}{\delta\rho(x)\delta\rho(x')}\int \frac{\delta C[\rho]}{\delta\rho(x'')}\frac{\delta A[\rho]}{\delta\rho(x'')}dx'' \ . \quad (30)$$

Eq.(30) is a ready-to-use, general formula to obtain a constrained second derivative that is in accordance with Eq.(9) and is applicable in stationary point analysis (see below) – for any constraint $C[\rho] = C$, not requiring one to take extra efforts to find a proper $\rho_C[\rho]$. Note further that Eq.(30) is applicable also when the functional variable $\rho(x)$ is a multiple variable; in this case, the functional derivative $\frac{\delta}{\delta\rho(x)}$ denotes a vector, while the second functional derivative $\frac{\delta^2}{\delta\rho(x)\delta\rho(x')}$ will be a matrix, and the multiplications of derivatives in Eq.(30) will be scalar product of vectors or matrix products.

A cornerstone of the appropriateness of the constrained second derivative Eq.(30) is as to whether it reduces to Eq.(8) for a constrained stationary point or not. Comparison of Eqs.(8) and (28) and use of the Euler-Lagrange equation $\frac{\delta A[\rho]}{\delta\rho(x)} = \mu\frac{\delta C[\rho]}{\delta\rho(x)}$ show that the only thing to prove is

$$\int \frac{\delta C[\rho]}{\delta\rho(x'')}\frac{\delta^2 \rho(x'')}{\delta_C\rho(x')\delta_C\rho(x)}dx'' + \int \frac{\delta^2 C[\rho]}{\delta\rho(x''')\delta\rho(x'')}\frac{\delta\rho(x''')}{\delta_C\rho(x')}\frac{\delta\rho(x'')}{\delta_C\rho(x)}dx'' = 0 \ . \quad (31)$$

But this holds by construction of $\rho_C[\rho]$ according to Eq.(25) ! Notice that the left side of Eq.(31) is nothing else than the constrained second derivative of $C[\rho]$, according to Eq.(28); that is, Eq.(31) can be written as

$$\frac{\delta^2 C[\rho]}{\delta_C \rho(x) \delta_C \rho(x')} = 0 \ . \tag{32}$$

Thus, we have Eqs.(20) and (32) as essential criteria for the constrained first and second derivatives, i.e. these constrained derivatives of the constraint must vanish. It is obvious from the derivation of Eqs.(18) and (25) that this must be true for higher order as well; that is, the constrained derivative of order $m$ must be constructed so that

$$\frac{\delta^n C[\rho]}{\delta_C \rho^n} = 0 \tag{33}$$

is satisfied for $n \leq m$. Note that originally, Eq.(33), for any $n$, was a trivial consequence of the construction of $\rho_C[\rho]$ (which fully satisfied the constraint for any $\rho(x)$) – or, in another view, $\rho_C[\rho]$ was constructed so that Eq.(33) be satisfied for any $n$.

By following the above procedure, we can derive constrained derivatives of any order. In third order, the proper $\rho_C[\rho]$ must satisfy

$$\iiint \frac{\delta^2 C[\rho_C]}{\delta \rho(x_1) \delta \rho(x_2) \delta \rho(x_3)} \frac{\delta \rho_C[\rho_C](x_1)}{\delta \rho(x)} \frac{\delta \rho_C[\rho_C](x_2)}{\delta \rho(x')} \frac{\delta \rho_C[\rho_C](x_3)}{\delta \rho(x'')} dx_1 dx_2 dx_3$$

$$+ \iint \frac{\delta^2 C[\rho_C]}{\delta \rho(x_1) \delta \rho(x_2)} \frac{\delta^2 \rho_C[\rho_C](x_1)}{\delta \rho(x'') \delta \rho(x')} \frac{\delta \rho_C[\rho_C](x_2)}{\delta \rho(x)} dx_1 dx_2 + \iint \frac{\delta^2 C[\rho_C]}{\delta \rho(x_1) \delta \rho(x_2)} \frac{\delta^2 \rho_C[\rho_C](x_1)}{\delta \rho(x'') \delta \rho(x)} \frac{\delta \rho_C[\rho_C](x_2)}{\delta \rho(x')} dx_1 dx_2$$

$$+ \iint \frac{\delta^2 C[\rho_C]}{\delta \rho(x_1) \delta \rho(x_2)} \frac{\delta^2 \rho_C[\rho_C](x_1)}{\delta \rho(x') \delta \rho(x)} \frac{\delta \rho_C[\rho_C](x_2)}{\delta \rho(x'')} dx_1 dx_2 + \int \frac{\delta C[\rho_C]}{\delta \rho(x''')} \frac{\delta^3 \rho_C[\rho_C](x''')}{\delta \rho(x'') \delta \rho(x') \delta \rho(x)} dx''' = 0 \ , \tag{34}$$

which can be inferred upon on the basis of

$$\frac{\delta^3 A[\rho]}{\delta_C \rho(x'') \delta_C \rho(x') \delta_C \rho(x)} = \iiint \frac{\delta^3 A[\rho]}{\delta \rho(x_1) \delta \rho(x_2) \delta \rho(x_3)} \frac{\delta \rho(x_1)}{\delta_C \rho(x)} \frac{\delta \rho(x_2)}{\delta_C \rho(x')} \frac{\delta \rho(x_3)}{\delta_C \rho(x'')} dx_1 dx_2 dx_3$$

$$+ \iint \frac{\delta^2 A[\rho]}{\delta \rho(x_1) \delta \rho(x_2)} \frac{\delta^2 \rho(x_1)}{\delta_C \rho(x'') \delta_C \rho(x')} \frac{\delta \rho(x_2)}{\delta_C \rho(x)} dx_1 dx_2 + \iint \frac{\delta^2 A[\rho]}{\delta \rho(x_1) \delta \rho(x_2)} \frac{\delta^2 \rho(x_1)}{\delta_C \rho(x'') \delta_C \rho(x)} \frac{\delta \rho(x_2)}{\delta_C \rho(x')} dx_1 dx_2$$

$$+ \iint \frac{\delta^2 A[\rho]}{\delta \rho(x_1) \delta \rho(x_2)} \frac{\delta^2 \rho(x_1)}{\delta_C \rho(x') \delta_C \rho(x)} \frac{\delta \rho(x_2)}{\delta_C \rho(x'')} dx_1 dx_2 + \int \frac{\delta A[\rho]}{\delta \rho(x''')} \frac{\delta^3 \rho(x''')}{\delta_C \rho(x'') \delta_C \rho(x') \delta_C \rho(x)} dx''' \ , \tag{35}$$

just as in the first- and second-order cases. Substitution of Eq.(26) for $\rho_C[\rho]$ in Eq.(34) shows that one needs to add a correction of Eq.(26) that accounts for the cancellation of a term $\dfrac{\delta C[\rho]}{\delta \rho(x)} \dfrac{\delta C[\rho]}{\delta \rho(x')} \dfrac{\delta C[\rho]}{\delta \rho(x'')} \dfrac{1}{Q^3} \left( 3 \int \left( \int \dfrac{\delta^2 C[\rho]}{\delta \rho(x_1) \delta \rho(x_3)} \dfrac{\delta C[\rho]}{\delta \rho(x_1)} dx_1 \int \dfrac{\delta^2 C[\rho]}{\delta \rho(x_3) \delta \rho(x_2)} \dfrac{\delta C[\rho]}{\delta \rho(x_2)} dx_2 \right) dx_3 \right.$

$$-\iiint \frac{\delta^2 C[\rho]}{\delta\rho(x_1)\delta\rho(x_2)\delta\rho(x_3)} \frac{\delta C[\rho]}{\delta\rho(x_1)} \frac{\delta C[\rho]}{\delta\rho(x_2)} \frac{\delta C[\rho]}{\delta\rho(x_3)} dx_1 dx_2 dx_3 \biggr).$$ This may be achieved by a modification

$$\rho_C[\rho] = \rho(x) - \frac{1}{Q[\rho]} \frac{\delta C[\rho]}{\delta\rho(x)} (C[\rho]-C) - \frac{1}{2!(Q[\rho])^2} \frac{1}{2} \frac{\delta Q[\rho]}{\delta\rho(x)} (C[\rho]-C)^2$$

$$-\frac{1}{3!(Q[\rho])^3} \frac{1}{2} \left( \int \frac{\delta^2 Q[\rho]}{\delta\rho(x)\delta\rho(x')} \frac{\delta C[\rho]}{\delta\rho(x')} dx' - 5\int \frac{\delta^2 C[\rho]}{\delta\rho(x)\delta\rho(x')} \frac{\delta Q[\rho]}{\delta\rho(x')} dx' \right)(C[\rho]-C)^3 \quad (36)$$

of Eq.(26), where Eq.(27) has been used to simplify notation. The corresponding constrained third derivative can then be obtained through Eq.(35), using Eqs.(21) and (29) and the third derivative of $\rho_C[\rho]$ (at $\rho_C(x)$), directly calculated from Eq.(36).

However, there is an ambiguity problem with the above-described origination of higher-order constrained derivatives. The factors of $(C[\rho]-C)^n$ in $\rho_C[\rho]$ are not uniquely determined by the requirements Eqs.(25) and (34) for $n=2,3$, or generally, Eq.(33). More concretely speaking, e.g. the factor of $(C[\rho]-C)^2$ in Eq.(26) has been determined on the basis that the terms between brackets in the fourth term of Eq.(30) should cancel each other for $A[\rho] = C[\rho]$; however, we may add to $\int \frac{\delta^2 C[\rho]}{\delta\rho(x)\delta\rho(x')} \frac{\delta C[\rho]}{\delta\rho(x')} dx'$ any $\sigma(x) \Big/ \frac{\delta C[\rho]}{\delta\rho(x)}$ with $\sigma(x)$ integrating to zero, and the term $\iint \frac{\delta C[\rho]}{\delta\rho(x'')} \frac{\delta C[\rho]}{\delta\rho(x''')} \frac{\delta^2 A[\rho]}{\delta\rho(x'')\delta\rho(x''')} dx'' dx'''$ in the brackets will still be cancelled, for $A[\rho] = C[\rho]$, by the other term, coming from the factor of $(C[\rho]-C)^2$ in $\rho_C[\rho]$. This ambiguity does not have relevance as regards stationary point analyses, but it remains a question as to which concrete definition is the proper one for generally valid higher-order constrained derivatives.

To close this section, it is worth mentioning that $\rho_C[\rho]$ of Eqs.(14), (26) and (36) may be considered as a truncated series expansion of a $\rho_C[\rho]$, if exists for the given constraint, that fully satisfies the constraint. Consider the expansion of such a $\rho_C[\rho]$ in $C$ around $C = C[\rho]$:

$$\rho_C[\rho](x) = \rho(x) + \frac{\partial \rho_C[\rho](x)}{\partial C}\bigg|_{C=C[\rho]} (C - C[\rho]) + \frac{1}{2!} \frac{\partial^2 \rho_C[\rho](x)}{\partial C^2}\bigg|_{C=C[\rho]} (C - C[\rho])^2 + \dots, \quad (37)$$

noting that $\rho_{C[\rho]}[\rho](x) = \rho(x)$. We are then looking for a $\rho_C[\rho]$ for which $\frac{\partial \rho_C[\rho](x)}{\partial C}\bigg|_{C=C[\rho]} = \frac{1}{Q[\rho]} \frac{\delta C[\rho]}{\delta\rho(x)}$, etc. Such a $\rho_C[\rho]$ won't exist in a closed form in general.

For the simple constraint $\int \rho^2(x)\,dx = C$, however, we have such a $\rho_C[\rho]$ – namely, $\rho_C[\rho] = \sqrt{\dfrac{C}{\int \rho^2(x')\,dx'}}\,\rho(x)$, which indeed fully satisfies the constraint and gives Eq.(9) as constrained first derivative; therefore, we might expect this $\rho_C[\rho]$ to deliver (via Eq.(2)) *the desired unique higher-order constrained derivatives*, too, for the above constraint.

## III. Higher-order constrained derivatives obtained through successive constrained differentiations

Here, we will show that second-order constrained derivatives obtained through two successive constrained differentiations are also a proper tool for stationary point analyses (which fact was not recognized in [3]), which raises the possibility that actually, there is no need for a separate definition of higher-order constrained derivatives, but successive constrained differentiations provide proper higher-order constrained derivatives for a general application.

Consider a $\rho_C[\rho]$ that satisfies the given constraint to first order at least. The constrained first derivative emerging from this $\rho_C[\rho]$ may be written as

$$\frac{\delta A[\rho]}{\delta_C \rho(x)} = \int \frac{\delta \rho(x'')}{\delta_C \rho(x)} \frac{\delta A[\rho]}{\delta \rho(x'')}\,dx'' \ . \tag{38}$$

That is, the operator $\int dx'' \dfrac{\delta \rho(x'')}{\delta_C \rho(x)}$ projects $\dfrac{\delta A[\rho]}{\delta \rho(x'')}$ into $\dfrac{\delta A[\rho]}{\delta_C \rho(x)}$ – actually, any choice of Eq.(5) will yield a projection; the only criterion is that $u(x)$ integrate to one. Taking the constrained derivative of Eq.(38) gives

$$\frac{\delta \delta A[\rho]}{\delta_C \rho(x') \delta_C \rho(x)} = \iint \frac{\delta \rho(x'')}{\delta_C \rho(x)} \frac{\delta \rho(x''')}{\delta_C \rho(x')} \frac{\delta A[\rho]}{\delta \rho(x'') \delta \rho(x''')}\,dx''dx''' + \int \frac{\delta \delta \rho(x'')}{\delta_C \rho(x') \delta_C \rho(x)} \frac{\delta A[\rho]}{\delta \rho(x'')}\,dx'' \ , \tag{39}$$

where the notation $\delta\delta$ in the numerator signifies that the given second-order derivative is obtained through two successive differentiations. Since for $A[\rho] = C[\rho]$, the constrained first derivative, Eq.(5), vanishes, the constrained derivative of Eq.(5) will vanish too for $A[\rho] = C[\rho]$,

$$\frac{\delta \delta C[\rho]}{\delta_C \rho(x') \delta_C \rho(x)} = 0 \ , \tag{40}$$

giving

$$\iint \frac{\delta \rho(x'')}{\delta_C \rho(x)} \frac{\delta \rho(x''')}{\delta_C \rho(x')} \frac{\delta C[\rho]}{\delta \rho(x'')\delta \rho(x''')} dx'' dx''' + \int \frac{\delta \delta \rho(x'')}{\delta_C \rho(x')\delta_C \rho(x)} \frac{\delta C[\rho]}{\delta \rho(x'')} dx'' = 0 \ . \qquad (41)$$

Our aim is to prove that for having a local minimum of $A[\rho]$ under a given constraint, a necessary condition is

$$\iint \frac{\delta \delta A[\rho]}{\delta_C \rho(x)\delta_C \rho(x')} \Delta\rho(x)\Delta\rho(x') dx dx' \quad \text{be non-negative for any } \Delta\rho(x), \quad (42a)$$

and a sufficient condition is

$$\iint \frac{\delta \delta A[\rho]}{\delta_C \rho(x)\delta_C \rho(x')} \Delta\rho(x)\Delta\rho(x') dx dx' \quad \text{be strongly positive for any}^* \ \Delta\rho(x) \neq 0, (42b)$$

these being the well-known necessary and sufficient conditions for a local minimum in the unconstrained case (with $\dfrac{\delta^2 A[\rho]}{\delta \rho(x)\delta \rho(x')}$ in the place of $\dfrac{\delta \delta A[\rho]}{\delta_C \rho(x)\delta_C \rho(x')}$, of course). However, the star in Eq.(42b) denotes an important restriction on "any"; namely,

$$\Delta\rho(x) \neq u(x) \bigg/ \frac{\delta C[\rho]}{\delta \rho(x)} \ . \qquad (43)$$

Our starting point, just as in [3], is the following theorem (based on Ljusternik's theorems) [1]: To have a local minimum of $A[\rho]$ under a given constraint,

$$\iint \left( \frac{\delta^2 A[\rho]}{\delta \rho(x)\delta \rho(x')} - \mu \frac{\delta^2 C[\rho]}{\delta \rho(x)\delta \rho(x')} \right) \Delta_{C'}\rho(x)\Delta_{C'}\rho(x') dx dx' \quad \text{be non-negative for any } \Delta_{C'}\rho(x), \qquad (44a)$$

is a necessary condition, while

$$\iint \left( \frac{\delta^2 A[\rho]}{\delta \rho(x)\delta \rho(x')} - \mu \frac{\delta^2 C[\rho]}{\delta \rho(x)\delta \rho(x')} \right) \Delta_{C'}\rho(x)\Delta_{C'}\rho(x') dx dx' \quad \text{be strongly positive for any } \Delta_{C'}\rho(x) \neq 0, (44b)$$

gives a sufficient condition. Now, the key idea is to write a general $\Delta_{C'}\rho(x)$ as

$$\Delta_{C'}\rho(x) = \int \frac{\delta \rho(x)}{\delta_C \rho(x')} \Delta\rho(x') dx' \ , \qquad (45)$$

which can be done since the right-hand side of Eq.(45) always satisfies Eq.(15) and any $\Delta_{C'}\rho(x)$ can be written as Eq.(45). To prove these two statements, write out $\dfrac{\delta \rho(x)}{\delta_C \rho(x')}$ explicitly,

$$\frac{\delta \rho(x)}{\delta_C \rho(x')} = \delta(x-x') - u(x) \frac{\delta C[\rho]}{\delta \rho(x')} \bigg/ \frac{\delta C[\rho]}{\delta \rho(x)} \ . \qquad (46)$$

Then, the first statement is trivially proved, since integrating in $x$ Eq.(46) multiplied by $\dfrac{\delta C[\rho]}{\delta \rho(x)}$ yields zero, while the second statement is proved on the basis of the fact that for $\Delta \rho(x) = \Delta_{C'} \rho(x)$, Eq.(45) gives an identity – which also trivially follows from Eq.(46), with the use of Eq.(15). With Eq.(45), $\iint \left( \dfrac{\delta^2 A[\rho]}{\delta \rho(x) \delta \rho(x')} - \mu \dfrac{\delta^2 C[\rho]}{\delta \rho(x) \delta \rho(x')} \right) \Delta_C \rho(x) \Delta_C \rho(x') dx dx'$ in Eq.(44) can be replaced by

$$\iint \left\{ \iint \dfrac{\delta \rho(x'')}{\delta_C \rho(x)} \left( \dfrac{\delta^2 A[\rho]}{\delta \rho(x'') \delta \rho(x''')} - \mu \dfrac{\delta^2 C[\rho]}{\delta \rho(x'') \delta \rho(x''')} \right) \dfrac{\delta \rho(x''')}{\delta_C \rho(x')} dx'' dx''' \right\} \Delta \rho(x) \Delta \rho(x') dx dx' \; ,$$

and the two conditions can be stated for unconstrained variations $\Delta \rho(x)$, with the exception in the "sufficient" case that the corresponding $\Delta_{C'} \rho(x)$, delivered by Eq.(45), cannot be zero, which with the use of Eq.(46), gives Eq.(43). Now, utilizing Eq.(41), $\mu \dfrac{\delta C[\rho]}{\delta \rho(x)} = \dfrac{\delta A[\rho]}{\delta \rho(x)}$, and Eq.(39), the above expression may be rewritten as $\iint \dfrac{\delta \delta A[\rho]}{\delta_C \rho(x) \delta_C \rho(x')} \Delta \rho(x) \Delta \rho(x') dx dx'$ – so Eq.(42) is proved. (For the treatment of strong positivity in this kind of proofs, see [5].)

The conditions Eqs.(42) may be transformed into conditions concerning the eigenvalues of $\dfrac{\delta \delta A[\rho]}{\delta_C \rho(x) \delta_C \rho(x')}$, similarly to the case of Eq.(1) [3,5]. If $A[\rho]$ has a local minimum at some $\rho(x)$ under a given constraint, all the eigenvalues of $\dfrac{\delta \delta A[\rho]}{\delta_C \rho(x) \delta_C \rho(x')}$ for that $\rho(x)$ must be non-negative, while if all eigenvalues are positive (with zero not being an accumulation point of them) except for the zero eigenvalue of $\Delta \rho(x) = u(x) \Big/ \dfrac{\delta C[\rho]}{\delta \rho(x)}$, $A[\rho]$ has a strict local minimum at the given point (reverse for a local maximum). Note that the above exception will always be an eigenfunction of the constrained second derivative at a stationary point, with eigenvalue zero; this is true for constrained second derivatives obtained by Eq.(1), too, due to Eq.(8), with Eq.(46). On the basis of this, actually the constrained second derivative test carried out in [5] is conclusive even for a (non-degenerate) ground state, since the zero eigenvalue there must be disregarded.

Thus, we have that we may apply the constrained differentiation formula Eq.(9) twice and the obtained second-order derivative will be a proper tool in a second-order stationary point analysis. It may not be too bold to expect this to be true for higher order as well. We note that the series expansion Eq.(3) can also be written with higher-order constrained derivatives obtained through successive constrained differentiations. For its general use, it is worth giving the above second derivative explicitly:

$$\frac{\delta\delta A[\rho]}{\delta_C\rho(x')\delta_C\rho(x)} = \frac{\delta^2 A[\rho]}{\delta\rho(x')\delta\rho(x)} - \frac{1}{Q}\frac{\delta C[\rho]}{\delta\rho(x)}\left(\int\frac{\delta C[\rho]}{\delta\rho(x'')}\frac{\delta^2 A[\rho]}{\delta\rho(x')\delta\rho(x'')}dx'' + \int\frac{\delta^2 C[\rho]}{\delta\rho(x')\delta\rho(x'')}\frac{\delta A[\rho]}{\delta\rho(x'')}dx''\right.$$

$$\left. -\frac{2}{Q}\int\frac{\delta C[\rho]}{\delta\rho(x'')}\frac{\delta^2 C[\rho]}{\delta\rho(x')\delta\rho(x'')}dx''\int\frac{\delta C[\rho]}{\delta\rho(x''')}\frac{\delta A[\rho]}{\delta\rho(x''')}dx'''\right)$$

$$-\frac{1}{Q}\frac{\delta C[\rho]}{\delta\rho(x')}\left(\int\frac{\delta C[\rho]}{\delta\rho(x'')}\frac{\delta^2 A[\rho]}{\delta\rho(x)\delta\rho(x'')}dx'' - \frac{1}{Q}\int\frac{\delta C[\rho]}{\delta\rho(x'')}\frac{\delta^2 C[\rho]}{\delta\rho(x)\delta\rho(x'')}dx''\int\frac{\delta C[\rho]}{\delta\rho(x''')}\frac{\delta A[\rho]}{\delta\rho(x''')}dx'''\right)$$

$$+\frac{1}{Q^2}\frac{\delta C[\rho]}{\delta\rho(x)}\frac{\delta C[\rho]}{\delta\rho(x')}\left(\iint\frac{\delta C[\rho]}{\delta\rho(x'')}\frac{\delta C[\rho]}{\delta\rho(x''')}\frac{\delta^2 A[\rho]}{\delta\rho(x'')\delta\rho(x''')}dx''dx''' + \iint\frac{\delta^2 C[\rho]}{\delta\rho(x'')\delta\rho(x''')}\frac{\delta C[\rho]}{\delta\rho(x'')}\frac{\delta A[\rho]}{\delta\rho(x''')}dx''dx'''\right.$$

$$\left. -\frac{2}{Q}\iint\frac{\delta C[\rho]}{\delta\rho(x'')}\frac{\delta C[\rho]}{\delta\rho(x''')}\frac{\delta^2 C[\rho]}{\delta\rho(x'')\delta\rho(x''')}dx''dx'''\int\frac{\delta C[\rho]}{\delta\rho(x'''')}\frac{\delta A[\rho]}{\delta\rho(x'''')}dx''''\right)$$

$$-\frac{1}{Q}\frac{\delta^2 C[\rho]}{\delta\rho(x)\delta\rho(x')}\int\frac{\delta C[\rho]}{\delta\rho(x'')}\frac{\delta A[\rho]}{\delta\rho(x'')}dx'' \ . \tag{47}$$

To demonstrate how the above second derivative test works, we consider the finite-dimensional example that was considered in [9] and [10], to illustrate how the methods developed there for a stationary point analysis under constraint work. The "bordered Hessian" method of Hassell and Rees [9] is based on a modified second derivative matrix with dimensions increased by one (in the case of a single constraint), while the "restricted Hessian" method of Shutler [10] uses a modified second derivative matrix with dimensions decreased by one. The "constrained Hessian" method proposed here and in [3], in contrast, preserves the dimensions of the unconstrained, $n\times n$ Hessian, with $n$ denoting the number of the variables of the considered multi-variable function. The example is the following: Determine the characters of the stationary points of the function $a(x,y,z) = x^3 + y^3 + z^3$ under the constraint $x^{-1} + y^{-1} + z^{-1} = 1$. The stationary points of $a(x,y,z)$ under this constraint are $(3,3,3)$, $(1,1,-1)$, $(1,-1,1)$, and $(-1,1,1)$ [9]. The constrained derivative obtained with the use of Eq.(10) will be

$$\nabla_C a(x,y,z) = (3x^2, 3y^2, 3z^2) - \frac{-9}{x^{-4}+y^{-4}+z^{-4}}(-x^{-2},-y^{-2},-z^{-2}) \ , \tag{48}$$

while the constrained Hessian obtained by taking the constrained derivative of Eq.(48) arises as

$$\nabla_C\nabla_C a(x,y,z) = \begin{pmatrix} 6x+(12/Q)x^{-3}+(36/Q^3)(x^{-7}+y^{-7}+z^{-7})x^{-4}-(54/Q^2)x^{-7} & \dots & \dots \\ -(6/Q)y^{-1}x^{-2}-(36/Q^2)x^{-5}y^{-2}-(18/Q^2)y^{-5}x^{-2}+(36/Q^3)(x^{-7}+y^{-7}+z^{-7})x^{-2}y^{-2} & \dots & \dots \\ \dots & \dots & \dots \end{pmatrix},$$

$$\tag{49}$$

where the unfilled elements follow readily from the given ones by symmetry considerations, and $Q = |\nabla c|^2 = x^{-4} + y^{-4} + z^{-4}$. (Eq.(49) may also be deduced directly from the finite-dimensional form of Eq.(47).) For the stationary point (3,3,3), we have from Eq.(49)

$$\nabla_C \nabla_C a(3,3,3) = \begin{pmatrix} 24 & -12 & -12 \\ -12 & 24 & -12 \\ -12 & -12 & 24 \end{pmatrix}, \qquad (50a)$$

which has the eigenvalues 36, 36, and 0. For (1,1,-1), we get

$$\nabla_C \nabla_C a(1,1,-1) = \begin{pmatrix} 16/3 & -20/3 & 4/3 \\ -20/3 & 16/3 & 4/3 \\ 4/3 & 4/3 & -8/3 \end{pmatrix}, \qquad (50b)$$

with eigenvalues 12, –4, and 0, and for the other two stationary points, also the latter eigenvalues can be obtained. On the basis of these values, we can conclude that (3,3,3) is a local minimum point, while the other three points are saddle points (of index one), of $a(x, y, z)$ under the given constraint. The eigenvectors corresponding to the zero eigenvalues above are (1,1,1) (times an arbitrary constant) in all cases, and as can be checked readily, this vector is just $\nabla c(x, y, z)$ at the stationary points, times a constant; consequently, these zero eigenvalues should be disregarded in the analysis, on the basis of (the finite D version of) Eq.(43) with Eq.(7). (Note that in the case of a non-degenerate zero eigenvalue, this zero can always be disregarded.) On the other hand, in accordance with what has been shown in [5], the other eigenvectors vanish when multiplied by $\nabla c(x, y, z)$ at the corresponding stationary points. The above result of our illustrative stationary point analysis is in complete agreement with what was found in [9] and [10]. It is worth mentioning finally that for the constraint considered here, we can easily set up a $\rho_C[\rho]$ that fully satisfies the constraint; namely,

$$x_C(x,y,z) = \left(\frac{1}{x^{-1}+y^{-1}+z^{-1}}\right)^{-1} x, \ y_C(x,y,z) = \left(\frac{1}{x^{-1}+y^{-1}+z^{-1}}\right)^{-1} y, \text{ and } z_C(x,y,z) = \left(\frac{1}{x^{-1}+y^{-1}+z^{-1}}\right)^{-1} z. \text{ We}$$

can then use Eq.(1) to obtain a second derivative that may also be used in the analysis of the stationary points of $a(x, y, z)$. By carrying out the calculations, we get a constrained Hessian for (3,3,3) that is the same as Eq.(50a), while we get the constrained Hessian

$$\begin{pmatrix} 0 & -12 & -12 \\ -12 & 0 & -12 \\ -12 & -12 & -24 \end{pmatrix} \qquad (51)$$

for (1,1,–1), which are obtained from $\frac{\partial^2 a(x,y,x)}{\partial_c x^2} = 6x + 6(x^3 + y^3 + z^3)(x^{-3} + x^{-4}) - 18$ and

$\frac{\partial^2 a(x,y,x)}{\partial_c x \partial_c y} = -9x^2 y^{-2} - 9y^2 x^{-2} + 6x^{-2} y^{-2}(x^3 + y^3 + z^3)$, … (noting that the Hessian obtained by Eq.(1) will necessarily be symmetric [3]). Eq.(51) has the eigenvalues 12, –36, and 0, with corresponding eigenvectors (1,–1,0), (1,1,2), and (1,1,–1) (times constants), respectively. Then, the zero eigenvalue must be disregarded on the basis of Eq.(43), since Eq.(6) gives $(u_x, u_y, u_z) = (x^{-1}, y^{-1}, z^{-1})$; that is, the excluded $(\Delta x, \Delta y, \Delta z)$ is $-(x,y,z)$, taken at the stationary point (1,1,–1), which will be just the eigenvector of the zero eigenvalue times a constant. The zero eigenvalue of Eq.(50a) must be similarly disregarded here too, as $-(3,3,3)$ equals the eigenvalue (1,1,1) of Eq.(50a) (which corresponds to the zero eigenvalue) times a constant. We note again that a non-degenerate zero eigenvalue is always to be disregarded. Thus, we obtain that (3,3,3) is a local minimum point, while (1,1,–1) is a saddle point, in accordance with what has been obtained above. It is worth observing that the eigenvector (1,1,2) of Eq.(51) does not vanish when multiplied by $\nabla c(1,1,-1) = -(1,1,1)$ – the necessary first-order satisfaction of the constraint by the eigenvectors corresponding to nonzero eigenvalues being a property only of constrained second derivatives with Eq.(7).

## Appendix: Multiple constraints

In this Appendix, we give the two-constraint version of Eq.(9). The generalization of the kernel of the orthogonal projection $\int dx'' \frac{\delta \rho(x'')}{\delta_c \rho(x)}$, with $\frac{\delta \rho(x'')}{\delta_c \rho(x)}$ given by Eq.(21), can be given for two constraints as

$$\frac{\delta \rho(x)}{\delta_c \rho(x')} = \delta(x-x') - \frac{1}{|Q|}\left( Q_{22} \frac{\delta C_1[\rho]}{\delta \rho(x')} \frac{\delta C_1[\rho]}{\delta \rho(x)} - Q_{12} \frac{\delta C_1[\rho]}{\delta \rho(x')} \frac{\delta C_2[\rho]}{\delta \rho(x)} - Q_{12} \frac{\delta C_1[\rho]}{\delta \rho(x)} \frac{\delta C_2[\rho]}{\delta \rho(x')} + Q_{11} \frac{\delta C_2[\rho]}{\delta \rho(x')} \frac{\delta C_2[\rho]}{\delta \rho(x)} \right),$$

(A1)

where $Q_{ij}[\rho] = \int \frac{\delta C_i[\rho]}{\delta \rho(x)} \frac{\delta C_j[\rho]}{\delta \rho(x)} dx$ and $|Q[\rho]| = Q_{11}[\rho] Q_{22}[\rho] - Q_{12}^2[\rho]$. The constrained derivative of a functional $A[\rho]$ then emerges as

$$\frac{\delta A[\rho]}{\delta_{C_1,C_2} \rho(x)} = \frac{\delta A[\rho]}{\delta \rho(x)} - \frac{\delta C_1[\rho]}{\delta \rho(x)} \left( \frac{Q_{22}}{|Q|} \int \frac{\delta C_1[\rho]}{\delta \rho(x')} \frac{\delta A[\rho]}{\delta \rho(x')} dx' - \frac{Q_{12}}{|Q|} \int \frac{\delta C_2[\rho]}{\delta \rho(x')} \frac{\delta A[\rho]}{\delta \rho(x')} dx' \right)$$

$$-\frac{\delta C_2[\rho]}{\delta\rho(x)}\left(\frac{Q_{11}}{|Q|}\int\frac{\delta C_2[\rho]}{\delta\rho(x')}\frac{\delta A[\rho]}{\delta\rho(x')}dx' - \frac{Q_{12}}{|Q|}\int\frac{\delta C_1[\rho]}{\delta\rho(x')}\frac{\delta A[\rho]}{\delta\rho(x')}dx'\right). \quad (A2)$$

With Eq.(A2), the corresponding formulae for the fluid-dynamical model of Clarke [11] may also be obtained, with $\rho = (h, \varphi)$.